\documentclass{pasj00}
\RelativeFigurePath{./}
\begin{document}

\pagestyle{sample}

\SetRunningHead{Matsushita et al.}{Metals in the Coma Cluster}

\title{Distribution of Si, Fe, and Ni in
the Intracluster Medium of the Coma Cluster}

\author{
Kyoko \textsc{Matsushita}\altaffilmark{1}, 
Takuya \textsc{Sato}\altaffilmark{1}, 
Eri \textsc{Sakuma}\altaffilmark{1}, 
and 
Kosuke \textsc{Sato}\altaffilmark{1}.
}
\altaffiltext{1}{
Department of physics, Tokyo University of Science, 
1-3 Kagurazaka, Shinjuku-ku, Tokyo 162-8601 }
\email{matusita@rs.kagu.tus.ac.jp}
\Received{---}
\Accepted{---}
\Published{---}

\maketitle

\begin{abstract}
We studied  the distributions of Si, Fe, and Ni in the intracluster medium (ICM) of the
Coma cluster, one of the largest clusters in the nearby universe, 
 using XMM-Newton data up to 0.5 $r_{180}$ and Suzaku data 
of the central region up to 0.16 $r_{180}$.
Using the flux ratios of Ly$\alpha$ line of H-like Si  
 and 7.8 keV line blend to  
 K$\alpha$ line of He-like Fe, 
  the abundance ratios of Si to Fe and   Ni to Fe   of the ICM were derived  
using APEC model v2.0.1.
The Si/Fe ratio in the ICM of the Coma cluster shows no radial gradient.
The emission weighted averages of the Si/Fe ratio in the ICM 
within 0.0--0.2 $r_{180}$, 0.2--0.5 $r_{180}$, and 0.0--0.5 $r_{180}$
are 0.97 $\pm$ 0.11,  1.05 $\pm$ 0.36 and 0.99 $\pm$ 0.13, respectively,
in solar units using the solar abundance table by \citet{lodd2003}.
These values are  
 close to those of smaller clusters and groups of galaxies.
Using the Suzaku data of the central region,  
the derived Ni/Fe ratio
of the ICM is  0.6--1.5  in solar units,  
according to the same solar abundance table.
The systematic difference in the derived abundance ratios 
 by different plasma codes
are about 10\%.
Therefore, for the ICM in the Coma cluster, the abundance pattern of 
Si, Fe, and Ni is consistent with the same mixture of the yields of
 supernova (SN) II and SN Ia
in our Galaxy.
Within 0.5 $r_{180}$, the cumulative iron-mass-to-light ratio increases with 
radius, and  its radial profile  is similar to 
those of relaxed smaller clusters with cD galaxies at their center.
Using the observed Si/Fe ratio,
the cumulative metal-mass-to-light ratios at 0.5 $r_{180}$ are compared with
  theoretical expectations.
\end{abstract}

\section{Introduction}

An important clue to the  evolution of galaxies is the
metals in the intracluster medium (ICM).
Because the Fe--K lines are prominent in the spectra of the ICM,
the Fe abundance in the ICM has been studied in detail.
With ASCA observations,
\citet{Fukazawa2000} found that clusters with a sharp X-ray emission
centered on a cD galaxy commonly exhibit a central increment in the Fe abundance
of the ICM.
With Beppo-SAX observations, \citet{deGrandi2001} also found a difference in the
Fe abundance profiles between clusters with and without cool-cores.
With XMM-Newton observations, \citet{Johnson2011} found a similar trend among 
groups of galaxies.
Within cool cores of clusters and groups of
galaxies, the central Fe peak could have been mainly produced by type Ia
supernovae (SNe) in cD galaxies. In contrast, during cluster merging,
mixing of the ICM could destroy the cool cores and the central Fe peaks.

Since metals have been synthesized by SNe in
galaxies, the ratios of  metal mass in the ICM to the total light
from galaxies in clusters or groups, i.e., the metal-mass-to-light ratios,
are the key parameters in investigating the chemical evolution of the
ICM\@.
With ASCA observations,
the derived  ratios of Fe mass in the ICM to the total light from galaxies, 
iron-mass-to-light ratio (IMLR),  within a radius where the ICM density
falls below 3 $\times 10^{-4} ~\rm{cm^{-3}}$
 is nearly constant in rich clusters and decreases toward poorer systems \citep{Makishima2001}. In individual clusters, the IMLR is lower around the center \citep{Makishima2001}.
With Chandra, XMM, and Suzaku observations of groups and medium-size clusters, 
the lower IMLR  within a given over-density radius
  in some groups of galaxies have been confirmed 
\citep{Matsushita2007a,  Tokoi2008, Rasmussen2009,  Sato2009a, Sato2009b, kSato2010, Komiyama2009, Sakuma2011,Murakami2011}.

Since Fe is both synthesized in SN Ia and SN II, to constrain contributions from 
the two types of SN, we need measurements of abundances of various elements.
The ASCA satellite first studied the Si abundance in the ICM \citep{Fukazawa1998, Fukazawa2000, Finoguenov2000, Finoguenov2001}.
\citet{Fukazawa1998}
 reported that the Si/Fe ratio in the ICM increases with ICM temperature,
suggesting that the relative contribution of SN II increasing towards
massive clusters.  
\citet{Finoguenov2000} reported that the Si/Fe  ratio increases
with radius in several clusters.    
Using Chandra data of groups out to $r_{500}$, 
\citet{Rasmussen2007} found that SN II contribution increases with radius and completely dominates at  $r_{500}$.
 XMM-Newton and Suzaku observations also have been used to study
the Si/Fe ratio of the ICM in clusters and  groups of galaxies.
\citep{Matsushita2003, Tamura2004, Sanders2006, Werner2006,  dePlaa2007, Rasmussen2007,Matsushita2007a, Matsushita2007,  Sato2007a, Sato2007b, Sato2008, Tokoi2008, Komiyama2009, deGrandi2009, Simionescu2009, Sato2009a, Sato2009b, kSato2010, Sakuma2011, Murakami2011}.
 With Suzaku observations of clusters and groups with the ICM temperatures lower than
$\sim4 $ keV,   the derived values of Si abundance 
 are close to those of Fe  (the Si/Fe ratios are $\sim$0.8 in solar units) out to 0.2--0.3 $r_{180}$,
with a small scatter using the solar abundance table by \citet{lodd2003}.
In the core regions,
 the reported values of the Si/Fe ratios have a larger scatter
\citep{dePlaa2007, deGrandi2009}.
With XMM-Newton observations of nearby clusters, \citet{Tamura2004} 
found that the temperature dependence does not exist in the Si/Fe ratios.
For example, the Si/Fe ratio outside the cool core of the Perseus cluster
is 0.77 $\pm$ 0.25  in solar units using the same solar abundance table.
However, excluding cool core regions,
the error bars in the Si/Fe ratio of hotter clusters 
are very large.

With Beppo-SAX observations, \citet{deGrandi2002} reported that
cooling flow clusters show higher  Ni abundances in their cores 
than non-cooling flow clusters.
  \citet{B2005} analyzed  hundreds of clusters observed with ASCA and reported that
the average Ni abundance in hot clusters is about 1.3 solar, 
which is significantly higher than the Fe abundances in the ICM.
In the cool core of the Perseus cluster, the Ni/Fe ratio is consistent with the solar ratio
\citep{Churazov2004, tamura2009, Matsushita2011b},
whereas that of the Centaurus cluster is significantly higher \citep{Matsushita2011b}.
\citet{dePlaa2007} found that
the weighted average of the Ni/Fe ratios of core regions of nearby clusters was
1.4 $\pm$ 0.3 with respect to solar ratio by \citet{lodd2003}.
\citet{deGrandi2009}
 studied Si, Fe, and Ni abundances in the central regions 
of 26 local  clusters and discovered that the Ni/Fe ratio scatters significantly.
In our Galaxy, [Ni/Fe] of stars are $\sim$ 0, with no dependence of [Fe/H]
\citep{Edv1993, feltzing1998, Gratton2003}.
This result indicates that in our Galaxy both SN II and SN Ia synthesize
Ni in a similar manner as that of Fe.  
 In cool cores, the enhancement of Fe abundance indicate the metal production from 
cD galaxies (e.g. \cite{Hans2004}).
Therefore, the higher Ni/Fe ratios observed in some clusters
indicate that Ni synthesis in cD galaxies  differ from that in our Galaxy.
However, in some cases, Ni abundances are derived from the spectral fitting
including the Fe-L energy band. Then, Ni abundance is
derived from residuals  between data and the Fe-L model, 
which strongly depends on the plasma codes. 
With CCD detectors, the K$\alpha$ lines of He-like Ni and
K$\beta$ lines of He-like Fe are blended into a single bump at 7.8 keV,
 and  the derived Ni abundance  couples with  the effect of  resonant line scattering
\citep{Churazov2004, Matsushita2011b}.
Therefore,  for cases in which  the optical depth for 
the scattering is sufficiently small,
 spectral fittings of K lines of Ni give more reliable Ni abundances.

The Coma cluster ($z=0.0231$), also known as Abell~1656, is one of 
the largest clusters in the nearby universe.
The cluster does not have a strong cool-core in the center,
and the X-ray peak is not associated with  two dominant galaxies in the central region \citep{Vik94}.
With Chandra observations,  \citet{Vik01} found that
 these dominant galaxies retain their X-ray corona in the form of compact halo 
(a few kpc in size) with temperatures of 1--2 keV.
Using XMM-Newton data,
\citet{arnaud01} found a temperature drop within 1 arcmin 
from one of the dominant galaxies, NGC 4874.
However, they  reported that
 the projected temperature distribution 
of most of the core region is remarkably homogeneous.
Using  Suzaku data,  \citet{Sato2011} reported that
fittings  of 
the continuum spectra and the ratio of the 
Ly$\alpha$ line of H-like Fe and K$\alpha$ line 
of He-like Fe   implied    the same ICM temperatures.
Because line ratio is a steep function of temperature, 
this consistency supports the accuracy of temperature measurements 
using  Suzaku. 
\citet{Matsushita2011} and \citet{Sato2011} derived the Fe abundance profile 
in the ICM up to $\sim$ 0.5 $r_{180}$.
Within 0.2 $r_{180}$, the Fe abundance is flat at $\sim$0.4 solar,
according to the solar 
abundance table by \citet{lodd2003},
and further decreases with radius.
This flat Fe abundance   is significantly different 
from the peaked abundance profiles of cool-core clusters and  
indicates that
 the gases have been mixed well   in the core   during the 
past mergers associated with cluster growth.

In this paper, we study  Si/Fe and Ni/Fe ratios in
the ICM of the Coma cluster observed with XMM and Suzaku.
In addition, we derived  IMLR profiles of the
Coma cluster and compared the results with smaller clusters.
The paper is organized as follows; After the 
introduction, we present the observations in section 2,
 followed by the description of our data analysis and results in 
section 3.
In section 4, we discuss our results.

We used the Hubble constant, $H_{\rm 0} = 70$~km~s$^{-1}$~Mpc$^{-1}$\@.
The distance to the Coma cluster is $D_{\rm L}=101$~Mpc, and $1'$
corresponds to 28~kpc. 
The virial radius of the Coma cluster,
 $r_{180}=1.95~h_{100}^{-1}\sqrt{k \langle T \rangle/10~\rm{keV}}~\rm{Mpc}$
\citep{Evrard1996, Markevitch1998}, is 2.5~Mpc for the average temperature $k \langle T \rangle$=7.8 keV.
We used the solar abundance ratio by \citet{lodd2003},
in which the solar Si, Ni and  Fe abundances relative to H are
3.47$\times$10$^{-5}$,  1.66$\times$10$^{-6}$, and 2.95$\times$10$^{-5}$,
respectively,  by number.
Considering a difference in solar He abundance, 
the Fe abundance yielded by \citet{lodd2003}  
is 1.5 times higher than that using the photospheric value by \citet{angr}.
Using the table by \citet{lodd2003}, the Si/Fe and Ni/Fe ratios 
are factors of 1.55 and 1.48, respectively,  smaller than those from \citet{angr}.
Errors are quoted at 90\% confidence level for a single parameter.
The spectral analysis employed the XSPEC\_v12.7.0 package.

\section{Observation and Data Reduction}
\label{sec:obs}

\subsection{XMM observations}

\begin{figure}
\FigureFile(70mm, 60mm){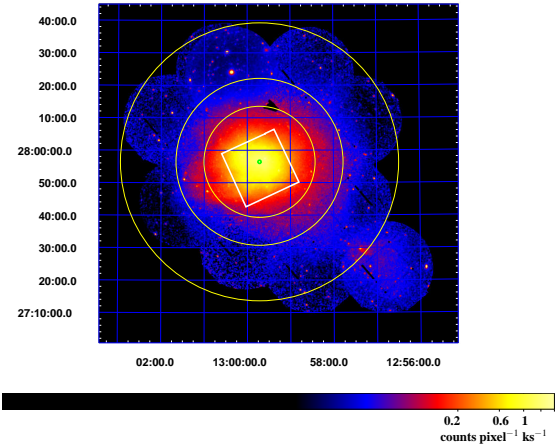}
\caption{Exposure-corrected combined MOS  image of the Coma cluster
(0.5--4.0 keV). The green circle corresponds to the X-ray peak,
and the white square indicates the field of view (FOV) of the Suzaku.
Yellow circles have radii of 0.2, 0.3, and 0.5 $r_{180}$. }
\label{fig:xmmimage}
\end{figure}

We analyzed 25 pointings of XMM-Newton observations of the Coma cluster
using   PN, MOS1, and MOS2 detectors. 
In this study, we used SASv10.0, but the
 details of observations, event selection, and  background subtraction
  are the same as those in \citet{Matsushita2011}.
The exposure corrected combined MOS image of the Coma cluster
within an energy range of 0.5--4.0 keV is shown in Figure \ref{fig:xmmimage}.
Spectra were  accumulated over  annular regions 
 centered on the X-ray peak of the cluster.
The spectra from MOS1 and MOS2 were added.

\subsection{Suzaku observation}

We analyzed Suzaku XIS \citep{koyama07} data of the central region of the Coma cluster,
whose Suzaku sequence number is 801097010,
observed in 2006 May with an exposure of 150 ks.
The XIS instrument consists of four  X-ray CCD 
(XIS0, 1, 2, and 3). XIS1 is a back-illuminated (BI) sensor, while 
XIS0, 2, and 3 are front-illuminated (FI) ones.
Figure 1 shows the field of view (FOV) of the XIS.
 We  used HEAsoft version 6.11\@, and CALDB 2011-02-10, but
the details of observations,  data reduction,  background subtractions
and response matrix are the same as those in  \citet{Sato2011}.
We extracted spectra over the FOV of the XIS, except for 
the regions  around calibration sources.

\section{Analysis and Results}

\subsection{Radial profile of  Si/Fe ratio}

\begin{figure}
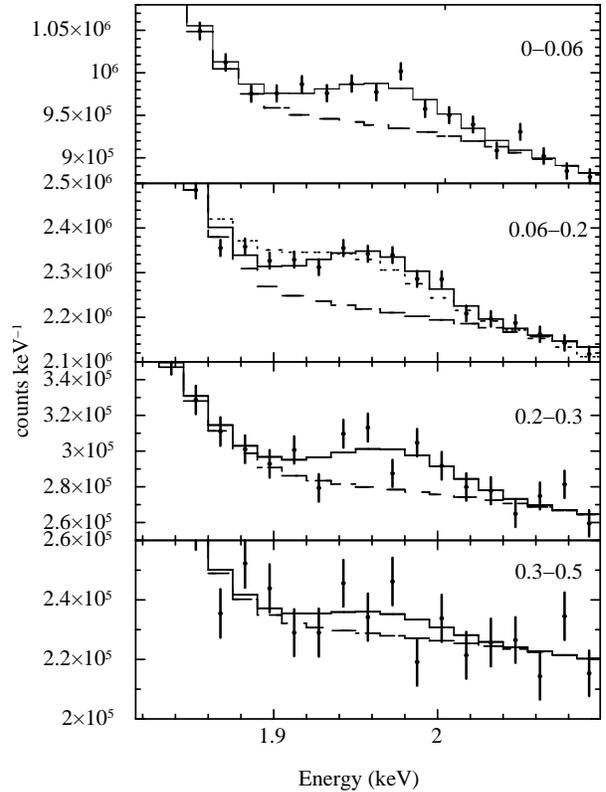

\FigureFile(80mm, 60mm){PASJ4137_figure2.ps}
\caption{ MOS spectra at 0.0-0.06, 0.06-0.2, 0.2--0.3, and
 0.3--0.5 $r_{180}$ (from top to bottom) fitted with an  vAPEC  v2.0.1    model with Si abundance = 0 solar (dashed line) and a Gaussian.
The dotted line at 0.06--0.2 $r_{180}$ corresponds to the best-fit single-temperature
APEC model fitted within an energy range of 1.8--2.1 keV,   where
 the best-fit Si abundance was 0.1 solar. 
}
\label{fig:sispec}
\end{figure}

With the CCD response 
  and a typical Si abundance,
the peak level    of the Ly$\alpha$ line of H-like Si of a 8 keV plasma
  in a spectrum  
  is only a few percent above  that of the continuum  
 and a small systematic uncertainty in the response matrix 
 can cause a large systematic uncertainty in the Si abundance.
We   first fitted    the MOS and PN spectra at 0.06--0.2 $r_{180}$ 
with a single-temperature vAPEC model v2.0.1 \citep{Smith2001}.
Using an energy range of 0.5--7.2 keV, the two kind of detectors gave
 temperatures and  Si abundances of  8.2--8.3 keV and 0.0 solar,
respectively, and there are discrepancies of a few percent
 between the data and model around the Ly$\alpha$ line of  H-like Si.
  Next,  we restricted the energy range to 1.8--2.1 keV,
and refitted the spectra. Then, the best-fit Si abundance derived from
MOS and PN were 0.1 and 0.5 solar, respectively.
In this case, reduced $\chi^2$ for the MOS spectrum was still greater than 2,
and the discrepancies between the data and model were around a few percent
at maximum (see Figure \ref{fig:sispec}).
In contrast,  $\chi^2$ for the PN spectrum became acceptable.
Furthermore, 
 the best-fit ICM temperatures, 5.4 keV and 11.7 keV, 
respectively, differed  significantly  from 
those derived from the wider energy range.
Using the older version of response matrices  by SAS-8.0 also gave significantly 
different Si abundances.
These results indicate   a minor difference  in the response files 
 around the Si line  
 caused a significant discrepancy in the derived Si abundance.
For the Suzaku XIS detectors,
 severe discrepancies at 1.7--1.9 keV and 2.1--2.2 
keV between the Suzaku FI and BI responses   were reported 
\citep{Sato2011}.

\begin{figure*}[t]
\FigureFile(80mm, 60mm){PASJ4137_figure3a.ps}
\FigureFile(80mm, 60mm){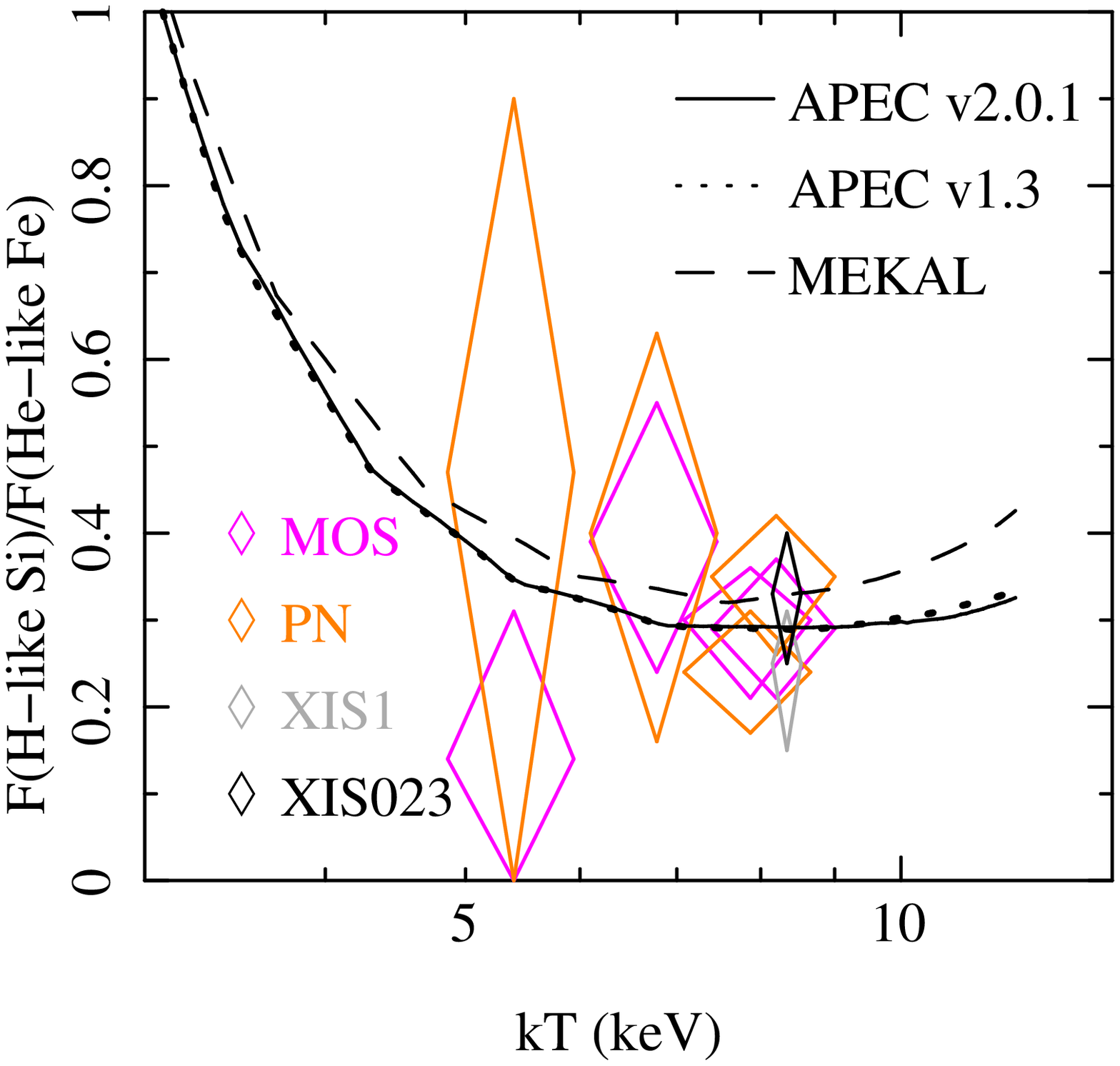}
\caption{
(left panel) The radial profiles of the Si/Fe ratio in solar units
derived from spectral fitting with vAPEC model v2.0.1 around the Si line (closed circles with 
error bars) using MOS and PN (red) and XIS detectors (blue).
Those derived from the line ratio of Ly$\alpha$ of H-like Si to K$\alpha$ of He-like Fe (diamonds) using MOS (magenta), PN (orange), XIS-FI (black), and XIS-BI (gray)
 are also plotted.
 Here, we used vAPEC  v2.0.1 model.  
(right panel)
Flux ratios of Ly $\alpha$ line of H-like
Si to K$\alpha$ line of He-like Fe derived from MOS (magenta),  
PN (orange),  Suzaku XIS-FI (black) and XIS-BI (gray) plotted against plasma temperature
derived by \citet{Matsushita2011} and \citet{Sato2011}.
The temperatures and line ratios were derived within the radial range
listed in Table 2.
We added a 10\% systematic uncertainty in plasma temperature derived with XMM-Newton
data \citep{Matsushita2011}.
The solid, dotted, and dashed lines indicate  theoretical ratios with a plasma
of solar Si/Fe ratio according to APEC v2.0.1, APEC v1.3,  and MEKAL 
plasma codes, respectively.
 
}
\label{fig:sifeprofile}
\end{figure*}

We added a 1\% systematic error to the counts of each channel in the spectra of MOS 
  and PN,  
and simultaneously fitted MOS and  PN spectra within energy ranges of 
1.8--2.1 keV and 1.8--2.2 keV, respectively,
with a single-temperature vAPEC v2.0.1 model.
We also fitted the XIS FI and BI spectra simultaneously within an energy range
of 1.92--2.1 keV in the same way.
Here, the Si abundance was allowed to vary and 
those of the other metals  were fixed to the best-fit Fe abundance
derived in \citet{Matsushita2011}.
Temperature was fixed to  the best-fit value derived in \citet{Matsushita2011},
since the equivalent width of the Si line at fixed Si abundance
 depends on the plasma temperature.
 Above 1.8 keV,  an systematic uncertainty caused by
 strong instrumental  fluorescence line at $\sim$ 1.7 keV of the MOS detector
does not affect the derived Si abundance.  
Table \ref{tab:siapec} shows  $\chi^2$ and  the derived Si/Fe ratios using
the Fe abundances derived in \citet{Matsushita2011}.
The $\chi^2$ became acceptable.
As shown in the left panel in Figure \ref{fig:sifeprofile},
 the derived  Si/Fe ratios are consistent with the 
solar ratio and show no radial gradient.
The XIS spectra within 0.16 $r_{180}$
 gave a smaller Si/Fe ratio of 0.7 $\pm$ 0.2, in units of the solar ratio.
The emission weighted averages of the 
 Si/Fe  ratio with MOS and PN  within 
 0.0--0.2 $r_{180}$, 0.2--0.5 $r_{180}$, 0.0--0.5 $r_{180}$
 are 0.94 $\pm$ 0.16, 1.09 $\pm$ 0.37 and 0.98 $\pm$ 0.16,
respectively,  in units of the solar ratio.

\begin{table}[t]
 \caption{Results of spectra fits 
 around the Ly$\alpha$-line of H-like Si with a vAPEC v2.0.1  model. }
\label{tab:siapec}
\begin{center}
\begin{tabular}{llrrcccc} 
\hline
radius & detector & Si/Fe  & $\chi^2/d.o.f$  \\
($r_{180}$) &  & (solar ratio) & \\
\hline
0.0--0.06 & MOS, PN  &1.00 $\pm$ 0.23    & 28/35  \\
0.06--0.2 & MOS, PN  &0.89 $\pm$ 0.21    & 34/35 \\
0.2--0.3  & MOS, PN  & 1.23 $\pm$ 0.49   & 36/35  \\
0.3--0.5  & MOS, PN  & 0.79 $\pm$ 0.53   & 39/35  \\
0.0-0.16  & XIS0123 & 0.70 $\pm$ 0.20   &14/19    \\
\hline
\end{tabular}
\end{center}
\end{table}

With the APEC model v2.0.1, the ratio of Ly$\alpha$ line of H-like Si and K$\alpha$ line of He-like Fe  is, at fixed Si/Fe ratio,
 constant to within 10\% across the $kT$=6--12 keV temperature range
(the right panel of Figure \ref{fig:sifeprofile}).
The weak temperature dependence of the line ratio can minimize the effect of
uncertainties in the temperature structure of the ICM.
Therefore,  we also derived the line flux of these two lines  
and converted their ratios  to  abundance ratios of Si to Fe
using the theoretical expectations.
To derive line strengths of the Ly$\alpha$ of  H-like Si, we
 fitted spectra   of MOS, PN, and XIS-BI and XIS-FI detectors
 within an energy range of 1.8--2.1 keV,
1.8--2.2 keV, 1.92--2.1 keV, and 1.92--2.1 keV, respectively,
with a  sum of vAPEC model and a Gaussian for the Ly $\alpha$ line of H-like Si.
Considering differences in  energy resolution, we used a wider energy range for
PN.
Here, we used spectra without systematic error to the counts of each channel.
The Si abundance of the vAPEC model was fixed at 0, and 
the abundances of the other metals were fixed to the best-fit Fe abundance
derived in \citet{Matsushita2011}.
Because of  the high ICM temperature of the Coma cluster, 
the K$\alpha$ line of He-like Si is negligible.
To derive line strengths of Fe, we also fitted these spectra within an energy range
of 6.0--7.2 keV with a bremsstrahlung and two Gaussians for He-like and H-like
Fe lines.

\begin{table}[t]
\caption{Line ratio of the Ly$\alpha$ of H-like Si and K$\alpha$ of He-like Fe and
the Si/Fe ratio}
\label{tab:sigauss}
\begin{center}
\begin{tabular}{llrrrcrrl} 
\hline
radius & detector  & $F_{\rm Si}/F_{\rm Fe}$ \footnotemark[$\ast$]& Si/Fe  \footnotemark[$\dagger$] &$\chi^2/d.o.f$\footnotemark[$\ddagger$]\\
($r_{180}$) &  & & (solar ratio) &\\
\hline
 0.0-0.06 & MOS     & 0.29$^{+0.08}_{-0.08}$ &1.00$^{+0.28}_{-0.28}$         & 18.1/15 \\
0.06-0.2 & MOS      & 0.30$^{+0.06}_{-0.09}$ &1.03$^{+0.22}_{-0.31}$   &  15.1/15\\
0.2-0.3  & MOS      & 0.39$^{+0.16}_{-0.15}$ &1.30$^{+0.58}_{-0.56}$           & 20.4/15 \\
0.3-0.5  & MOS      & 0.14$^{+0.17}_{-0.14}$ &0.39$^{+0.55}_{-0.39}$   & 18.8/15 \\
0.0-0.06 & PN       & 0.35$^{+0.07}_{-0.09}$ &1.20$^{+0.24}_{-0.32}$   & 13.7/11 \\
0.06-0.2 & PN       & 0.24$^{+0.07}_{-0.07}$ &0.82$^{+0.25}_{-0.24}$   & 7.0/11 \\
0.2-0.3  & PN       & 0.40$^{+0.23}_{-0.24}$ &1.33$^{+0.82}_{-0.84}$           &  8.6/11\\
0.3-0.5  & PN       & 0.47$^{+0.43}_{-0.47}$ &1.30$^{+1.42}_{-1.30}$   & 9.2/11 \\
0.0-0.16  & XIS1     & 0.25$^{+0.06}_{-0.10}$ & 0.86$^{+0.21}_{-0.35}$  & 4.5/7   \\
0.0-0.16  & XIS023    & 0.33$^{+0.07}_{-0.08}$ & 1.14$^{+0.24}_{-0.28}$  &8.8/7    \\%
\hline
\end{tabular}
\end{center}
 \footnotemark[$\ast$]
Ratio of flux in units of photons cm$^{-2} \rm{s}^{-1}$ of Ly$\alpha$ line of H-like Si and K$\alpha$ line of He-like Fe.\\
 \footnotemark[$\dagger$]  Si/Fe abundance ratio in units of solar ratio derived from the line ratio   using the theoretical 
expectation from APEC v2.0.1.
 \\
\footnotemark[$\ddagger$] $\chi^2$ and degrees of freedom of the spectral fitting around
the Ly$\alpha$ line of H-like Si.
\end{table}

Table \ref{tab:sigauss} summarizes the results of the line ratio.
The values of  $\chi^2$ are reasonable. 
As shown in Figure \ref{fig:sispec},
  the Ly$\alpha$ line of the H-like Si is clearly seen in the spectra,
and well reproduced with the model.
The MOS, PN, and XIS detectors  gave almost the same line ratios
of the Ly$\alpha$ of H-like Si to the K$\alpha$ line of He-like Fe
(Table \ref{tab:sigauss} and the right panel of Figure \ref{fig:sifeprofile}).
The derived line ratios are consistent with the theoretical expectations
by the APEC v2.0.1,  v1.3 and MEKAL\citep{mewe85, mewe86, kaastra, liedahl}   codes  with a solar Si/Fe ratio.
 Using APEC v2.0.1 code, 
we converted the derived line ratios to the abundance ratio of Si/Fe
  using the temperatures derived by
\citet{Matsushita2011}, since \citet{Sato2011} found that
 the single-temperature model fits the Suzaku spectra of the Coma cluster  
fairly well.  
Here, we allowed a 10\% systematic error in the 
temperature for the XMM data (see details in \cite{Matsushita2011}).
The systematic effect by the plasma code on the Si/Fe ratio
is expected to be insignificant 
because 
 the MEKAL and  APEC v2.0.1  codes  gave similar line ratios within 10 \% at 8 keV
  and APEC v1.3 and APEC v2.0.1 gave almost the same line ratio
 (Figure \ref{fig:sifeprofile}).
Therefore, 
the systematic difference in the derived Si/Fe ratio by different plasma codes
is about 10\%, which is smaller than the statistical errors.

The left panel of Figure \ref{fig:sifeprofile} also  
shows the radial profile of the  Si/Fe ratio derived from 
the line ratio.
The derived abundance ratios from MOS and PN agree very well with those derived 
from spectral fitting with APEC model around the Ly$\alpha$ line of H-like Si.
Using  MOS, PN and XIS detectors, 
the emission weighted  average of  the
 Si/Fe abundance ratio from the line ratio within 
 0.0--0.2 $r_{180}$, 0.2--0.5 $r_{180}$,  and  0.0--0.5 $r_{180}$  are
 0.97 $\pm$ 0.11, 1.05 $\pm$ 0.36, and 0.99 $\pm$ 0.13, respectively, in units of the solar ratio.
These values are consistent with those derived from the spectral fitting
around the Si lines within 10 \%.

\subsection{Ni/Fe ratio of the central region}

\begin{figure}[t]
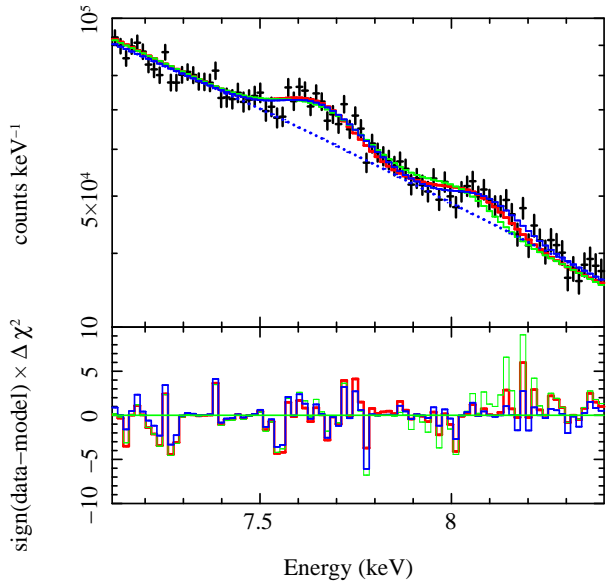

\FigureFile(80mm, 60mm){PASJ4137_figure4.ps}
\caption{  The spectrum of the XIS-FI of the central field of Suzaku fitted with a 
vAPEC v2.0.1 model (red),  vMEKAL model (green)
and bremsstrahlung and Gaussians (blue) within an energy range of 7.1--8.4 keV.
The blue dotted line corresponds to the contribution of the continuum of the Gaussian fit.
The bottom panel shows the contributions to the $\chi^2$.
}
\label{fig:specni}
\end{figure}

To derive Ni abundance, we used only Suzaku spectra of the central field
(0--0.16 $r_{180}$) because of its lower level of background.
 We added the spectra of the FI detectors, XIS0, XIS2, and XIS3.
We fitted background-subtracted
Suzaku spectra of the FI and BI detectors   simultaneously
 within an energy range of   3.0--8.5    keV
with a vAPEC plasma model v2.0.1.
Here, elemental abundances except for Fe and Ni were fixed at 0.4 solar,
since  Fe abundance of this region derived with Suzaku \citep{Sato2011} and XMM \citep{Matsushita2011} is nearly constant at this value.
The derived ICM temperature, the Fe and Ni abundances are 8.4 $\pm$ 0.05 keV, 
0.39 $\pm$ 0.01 solar, and 0.66 $\pm$ 0.10 solar,   respectively.
The derived ICM temperature and Fe abundance are consistent with those
derived by \citet{Sato2011} and \citet{Matsushita2011}.
However, the derived $\chi^2$, 719 for 560 degrees of freedom, was not acceptable,
considering only statistical errors.
Therefore, we restricted energy range to 7.1--8.4 keV and fitted the spectra again
with a vAPEC plasma model v2.0.1.
Here, 
the temperature and Fe abundance were fixed at  8.4 keV and 0.39 solar, respectively.
The results are shown in Figure \ref{fig:specni} and Table \ref{tab:nifit}.
The spectra were better fitted with the model, with $\chi^2$=142 for 128 degrees of freedom.
The derived Ni abundance, $0.48 \pm 0.13$ solar, became smaller than
the previous fit using the wider energy range.
When we used the energy range to 7.3--8.4 keV, the Ni abundance became
0.43 $\pm$ 0.14 solar, with a better reduced $\chi^2$ (Table \ref{tab:nifit}).
We also fitted the same spectra using  the energy range of 7.1--8.4 keV in the same way with
 vAPEC v1.3 and vMEKAL plasma codes.
As shown in Table \ref{tab:nifit}, 
the vAPEC v1.3 code gave almost the same results with the vAPEC v2.0.1 code.
The vMEKAL code also  gave almost the same Ni abundance, although
 the fit with the vMEKAL model around 8.2-8.3 keV, at the line blend of
 K$\gamma$ line of He-like Fe and K$\beta$ line of H-like Fe, 
  is significantly worse than 
 those using vAPEC models (Figure \ref{fig:specni}).

\begin{table}[t]
\caption{ Results of spectral fitting of the Suzaku central field\footnotemark[$\ast$].
  }
\label{tab:nifit}
\begin{center}
\begin{tabular}{lcccccccc} 
\hline
model   &  energy range & Ni   & $\chi^2/d.o.f$ \\
   & (keV) & (solar) \\
\hline
vAPEC v2.0.1 &7.1--8.4 & 0.48 $\pm$ 0.13 & 142/128  \\
vAPEC v2.0.1 &7.3--8.4 & 0.43 $\pm$ 0.14 & 109/109  \\
vAPEC v1.3 & 7.1--8.4    & 0.50 $\pm$ 0.13 & 146/128 \\
vMEKAL   & 7.1--8.4  & 0.51 $\pm$ 0.12 & 167/128 \\
\hline
\end{tabular}
\end{center}
 \footnotemark[$\ast$]
The temperature and the Fe abundances
were fixed at 8.4 keV and 0.39 solar, respectively.
\end{table}

\begin{table}[t]
\caption{  Results of spectral fits with a bremsstrahlung and Gaussians  using an energy range of 7.1--8.4 keV. }
\label{niline}
\begin{center}
 \begin{tabular}[t]{ccccc}
\hline
   Gaussian center \footnotemark[$\ast$] &  F(line)/F(6.7) \footnotemark[$\dagger$]  \\
   (keV)   & \\
\hline  
7.84 $\pm$ 0.02 & 0.166 $\pm$ 0.026 \\
8.27 $\pm$ 0.02 & 0.090 $\pm$ 0.021 \\
$\chi^2/d.o.f$ & 127/121\\
\hline
 \end{tabular}
\end{center}
\footnotemark[$\ast$] The central energy at the rest-frame of the Coma cluster\\
 \footnotemark[$\dagger$] Ratio of flux in units of photons cm$^{-2} \rm{s}^{-1}$
of the 7.8 keV or 8.3 keV blends to K$\alpha$ line of He-like Fe.
\end{table}

As in section 3.1 for the Si/Fe ratio,  
we derived the flux ratio of the 7.8 keV line blend to K$\alpha$ line
of He-like Fe line and converted it to the Ni/Fe ratio.
We fitted the same Suzaku spectra of XIS-FI and XIS-BI within the energy
range of 7.1--8.4 keV with a bremsstrahlung and two Gaussians for
the line blends of the 7.8 keV and 8.3 keV.
The line widths of the Gaussians are fixed to those derived from 
fitting of a mock XIS spectrum of 8.3 keV plasma with vAPEC v.2.0.1. 

\begin{figure}[t]
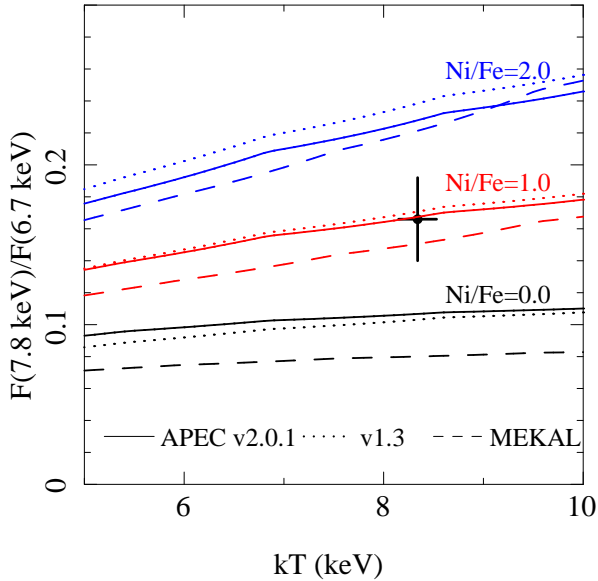

\FigureFile(80mm, 60mm){PASJ4137_figure5.ps}
\caption{ The ratio of flux of the 7.8 keV line blend and K$\alpha$ line of He-like Fe,
plotted against the plasma temperature.
The observed line ratio of the central field of Suzaku is plotted as the solid closed circle
with error bars. 
The solid, dotted, and dashed lines indicate theoretical line ratios according to
APEC v2.0.1, APEC v1.3, and MEKAL plasma codes, respectively, with plasma of Ni/Fe ratio of
0, 1, and 2, in units of the solar ratio.
}
\label{fig:niratio}
\end{figure}

The results are shown in Table \ref{fig:specni}.
The spectra were well represented with the model with 
$\chi^2=127$ for 121 degrees of freedom (Table \ref{niline}
and Figure \ref{fig:specni}).
Figure \ref{fig:niratio} shows the ratio of the flux of the 7.8 keV blend to 
that of the K$\alpha$ line of He-like Fe, plotted against the plasma temperature.
The theoretical  temperature dependence of the ratio 
and the discrepancy between the three codes are relatively small.
The derived ratio, 0.166 $\pm$ 0.026, agrees with the expectation of the APEC models
with the solar Ni/Fe ratio.
Converting the derived ratio to the abundance ratio,
the Ni/Fe ratio became 0.6--1.5, and 0.9--1.6, in units of solar ratio,
using the APEC codes and MEKAL code, respectively.

The worse $\chi^2$ with the MEKAL model reflects the discrepancy at the 8.3 keV line 
blend.
The derived central  energy of the Gaussian for the 8.3 keV blend, 8.27 $\pm$ 0.02 keV,
is consistent with  8.26 keV of the mock spectra of the APEC models.
However, it is significantly higher than the expected energy of 8.21 keV by the MEKAL model.
The derived flux ratio of the 8.3 keV blend and the 6.7 keV line is consistent with
the expectations by APEC and MEKAL codes.
Therefore,  the observed 8.3 keV line blend is more consistent with the
APEC codes rather than the MEKAL code.

\subsection{Iron-mass-to-light ratio}

Because  metals in the ICM   have been 
 synthesized in galaxies, the
metal-mass-to-light ratio is a useful measure for studying the chemical
evolution of clusters of galaxies.
To estimate the integrated Fe mass profile, we fitted 
the annular spectra of MOS within an energy range of 1.6--5.0 keV with a APEC model
to avoid uncertainties in the background.
Here, the temperature and Fe abundance were restricted within error bars
derived by \citet{Matsushita2011}.
The derived emissivity profile is well fitted with a single $\beta$-model.
The calculated  density profile from the best-fit $\beta$-model agrees well within 10\%  with that 
by \citet{briel1992} using ROSAT all sky survey data.
We used the Fe abundance profile by \citet{Matsushita2011} from XMM-Newton data
and derived integrated mass profile of Fe in the ICM.

Because the K-band luminosity of a galaxy correlates well with the stellar
mass, we calculated the luminosity profile of the K-band.
We collected K-band magnitudes of galaxies  in a
$6\times 6~{\rm deg}^2$ box centered on the center of Coma cluster from the Two
Micron All Sky Survey (2MASS).
We corrected foreground Galactic extinction of $A_\mathrm{K} = 0.003$
 \citep{Schlegel1998} from the NASA/IPAC Extragalactic Database.
The average surface brightness in the region at
 $100' < r < 170'$  was subtracted as the background.
Then, we deprojected the brightness profile assuming a spherical symmetry and
derived three-dimensional profile of K-band luminosity.

In addition,
we calculated the luminosity profile of the B-band to compare previous papers and
theoretical expectations.
We collected B-band magnitude of member galaxies of the Coma
cluster within 0.42 $r_{180}$ by
\citet{Michard2008} and within  $r_{180}$ by \citet{Doi}.
We corrected foreground Galactic extinction of $A_\mathrm{B} = 0.036$
 \citep{Schlegel1998} from the NASA/IPAC Extragalactic Database.
Then, we deprojected brightness profile as in the same manner as that for 
 the K-band luminosity profile and derived a three-dimensional profile of
B-band luminosity up to 0.5 $r_{180}$.
Since the data by \citet{Michard2008} is deeper than that by \citet{Doi},
we scaled the brightness profile by \citet{Doi} to match that by \citet{Michard2008}
and calculated the IMLR profile.

The derived IMLR profiles using K-band and B-band luminosity profiles
are plotted in Figure \ref{fig:imlr}.
Here, the error bars include only the abundance errors.
The cumulative profiles  of the IMLR increase with radius up to 0.5 $r_{180}$.
At a given radius,
the IMLR using B-band is a factor of 4--5 higher than that using the K-band.
The difference in the normalizations between the two bands 
is  consistent with a typical difference 
in the mass-to-light of stars in early-type galaxies \citep{Nagino2009}.
Therefore, the observed difference does not have a strong contribution from 
the different methods and data sets for the two bands.

\begin{figure}[t]
\FigureFile(80mm, 60mm){PASJ4137_figure6.ps}
\caption{
Radial profiles of integrated IMLR in the K-band (solid lines) 
and B-band (dashed lines) of the Coma cluster (black filled circles),
the Centaurus cluster ($kT=$3.9 keV; blue open circles; \cite{Sakuma2011}),
AWM 7 cluster ($kT=$3.5 keV; green open squares; \cite{Sato2008}),
Abell 262 cluster ($kT=$2.0 keV; magenta filled squares; \cite{Sato2009b}), 
 NGC 1550 group ($kT=$1.2 keV; red open triangles; \cite{kSato2010}),  
NGC 5044 group ($kT=$1.0 keV; orange diamonds; \cite{Komiyama2009}),
and the Fornax cluster ($kT=$1.3 keV; gray; \cite{Murakami2011}).
Here, the difference in the solar abundance tables are considered.
}
\label{fig:imlr}
\end{figure}

\section{Discussion}
\subsection{Contribution of SN Ia and SN II}

The emission weighted average of
the  Si/Fe ratio of the ICM within 0.5 $r_{180}$ of  the Coma cluster 
from the line ratio   
is 0.99 $\pm$ 0.13    in solar units.
This value is close to the  $\sim$ 0.8 in solar units
 of the Si/Fe ratios of the ICM of  clusters and of groups 
with ICM temperatures  smaller than several keV   \citep{Finoguenov2000, Rasmussen2007, Sato2007b, Sato2008, Sato2009b,deGrandi2009, Hayashi2009, Sato2011, Murakami2011, Sakuma2011}
and early-type galaxies \citep{Humphrey2006, Matsushita2007, Tawara2008, Hayashi2009}.
This result indicates that the Si/Fe ratio of the ICM does not depend on the
system mass because the Coma cluster is one of the largest clusters in 
the nearby universe.

Based on abundance ratios including Si and Fe, \citet{Finoguenov2000}, \citet{Humphrey2006}, \citet{dePlaa2007}, \citet{Rasmussen2007},  \citet{Sato2007b},
and \citet{deGrandi2009}
derived the contributions from SN Ia and SN II.
They found that using a classical deflagration model, W7 \citep{Iwamoto1999}, for the theoretical SN Ia yield, 
 over a half of Fe and a few tens \% of Si
 in the ICM were synthesized in SN Ia.
These papers also found that
the contribution of SN Ia to Si  strongly depends on the nucleosynthesis model:
with a delayed detonation (DD) model,  WDD1 \citep{Iwamoto1999}, 
about a half of Si comes from SN Ia.
 
We also calculated yield mixtures of nucleosynthesis models of SN II with
metallicity $=$ 0.02 by \citet{Nomoto2006} and SN Ia models of W7 and 
 WDDs  by \citet{Iwamoto1999}.
Using W7 and WDD3 models,  to explain the observed Si/Fe ratio 
of 0.99 $\pm$ 0.13  in solar units,
60\%--70\% (20\%--30\%)  of Fe (Si) were synthesized by SN Ia.
Using the WDD1 model
70\%--90\% (40\%--70\%) of Fe (Si) were originated from SN Ia.

In contrast to Fe and Si, O and Mg are predominantly synthesized in SNe II.
Abundance measurements covering a range of species from O to Fe are desired
 to constrain contributions and are 
therefore necessary to obtain unambiguous information on the history
of  formation  of massive stars. 
Suzaku enabled us to measure O and Mg abundances in 
clusters and groups of galaxies with ICM temperatures smaller than several keV
\citep{Matsushita2007a, Sato2007a, Tokoi2008, Sato2008, Sato2009, Komiyama2009, kSato2010, 
Sakuma2011, Murakami2011}.
The mixture of yields of W7 model and SN II gave better fits on the observed
 abundance pattern of O, Mg, Si, S and Fe in the ICM than that of WDD1 model and SN II
\citep{Sato2007a}.
With the next Japanese X-ray satellite ASTRO-H, we will be able to detect the O line 
from the  ICM around the X-ray peak of the  Coma cluster.

\subsection{Iron-mass-to-light ratio and constraints on  initial mass function}

\citet{Renzini2005} calculated 
 the ratios of the total amounts of O mass and Si mass synthesized
by SN II to present stellar luminosity in a cluster.
These ratios
are very sensitive functions of the slope of the initial mass function (IMF):
a decrease of the slope by 1 gives a factor of 20 higher O and Si masses
at fixed present-day total stellar luminosity. 
Adopting a Salpeter IMF
and O and Si yields by \citet{Woos1995},
 the expected values (ICM and stars) of 
the oxygen-mass-to-light ratio (OMLR) and silicon-mass-to-light ratio
(SMLR) are $\sim0.1~ M_\odot/L_{B,\odot}$
and $\sim 0.01~ M_\odot/L_{B,\odot}$, respectively.

We compared the observed IMLR of the Coma cluster within 
0.5 $r_{180}$ with the expected values by \citet{Renzini2005}
for a entire cluster.
At 0.5 $r_{180}$, the integrated IMLR for the ICM using B-band luminosity
is 0.016 $\pm ~0.001~$$M_{\odot}/L_{B, \odot}$.
Adopting the Si/Fe ratio of 0.99 $\pm$ 0.13 in solar units,
the integrated value of SMLR at 0.5 $r_{180}$
is $ 0.009 \pm 0.001 ~M_{\odot}/L_{B, \odot}$.
Adopting the W7 and WDD1 models for the SN Ia yield,
the SMLR within  this radius synthesized
by SN II are 0.006--0.008 $M_{\odot}/L_{B, \odot}$ and 0.002--0.005 $M_{\odot}/L_{B, \odot}$, respectively. 
Assuming the solar stellar metallicity in cluster galaxies,
the SMLR trapped in stars is 0.004 $M_{\odot}/L_{B, \odot}$
for stellar mass-to-light ratio of 5 \citep{Nagino2009}.
As a result, the sum of SMLR in the stars and the ICM within 0.5 $r_{180}$ 
is consistent 
 with the expected  SMLR of $\sim 0.01~ M_\odot/L_{B,\odot}$,
assuming a Salpeter IMF by \citet{Renzini2005}. 
The integrated IMLR profile of the Coma cluster increases with a radius 
up to 0.5 $r_{180}$. 
If the MLRs continue to increase radially beyond 0.5 $r_{180}$,a
 flatter, more top-heavy, IMF is necessary.
Therefore, to study the slope of the IMF in clusters of galaxies,  measurements
of MLRs at outer regions of these systems are desired.

\subsection{IMLR and dependence on the system mass}

Figure \ref{fig:imlr} compares the integrated profile of IMLR of the
Coma cluster with those of smaller systems observed with Suzaku,
the AWM 7 cluster ($kT\sim 3.5$ keV; \cite{Sato2008}),
the Centaurus cluster ($kT\sim 3.9$ keV; \cite{Sakuma2011}),
Abell 262 cluster ($kT\sim 2.0$ keV; \cite{Sato2009b}), 
the fossil group NGC 1550 ($kT\sim$1.2 keV; \cite{kSato2010}),
the NGC 5044 group ($kT\sim$1.0 keV; \cite{Komiyama2009}),
and the Fornax cluster ($kT\sim$1.3 keV; \cite{Murakami2011})
 using Suzaku data.
Here, differences in the  solar abundance tables were corrected.
The integrated IMLR of these clusters continues to increase with  a radius 
within the observed regions,
although the NGC 5044 group and the Fornax cluster show lower IMLR profiles,
and become flatter at 0.2--0.3 $r_{180}$.
Within 0.1 $r_{180}$, the derived IMLR profiles except the Fornax cluster
 agree with each other within a factor of two.
At 0.2 $r_{180}$, the integrated IMLR of the Coma clusters 
and poorer systems agree remarkably well, and
beyond 0.2 $r_{180}$, those of the Coma,  Abell 262 clusters and
the fossil group NGC 1550 system
agree very well.
At 0.35 $r_{180}$,  the AWM 7 cluster has a higher 
IMLR value than the Coma cluster.
The similarity of the IMLR profiles and Si/Fe ratios
among these clusters of galaxies indicates
these systems have universal metal enrichment histories.
The AWM 7 cluster was observed with Suzaku toward a direction 
to a  filament of the large-scale structure,
and the X-ray emission elongates with the same direction \citep{Sato2008}.
The anisotropy may cause a discrepancy in the IMLR profile.

The dependence of the IMLR on the system mass confirms
that derived from ASCA observations of groups and clusters of galaxies 
\citep{Makishima2001} 
  and also from Chandra observations \citep{Rasmussen2009}.
Groups of galaxies differ from richer systems in that their
IMLR  are systematically smaller than those in rich clusters.
Since the Fe abundances of systems plotted in Figure \ref{fig:imlr}
 at 0.1--0.5 $r_{180}$
are consistent within error bars \citep{Sato2009, kSato2010, Matsushita2011}, the difference in the IMLR 
reflects a difference in the ratio of gas  and stellar mass.
The stellar and gas mass fractions within $r_{500}$ depend on the total
system mass \citep{Lin2003, Lin2004, Vik06, Arnaud07, Sun09, Gio09}.
These studies  found that the
 stellar-to-total-mass ratios within $r_{500}$ of the groups are much 
larger than those in the clusters, whereas the gas mass fraction increases with
the system mass.
The observed higher stellar mass fraction and the lower gas mass fraction
 within $r_{500}$ in poor systems,
are occasionally interpreted as proof  that the
 star formation efficiency depends on the system mass.
However, as shown in Figure \ref{fig:imlr}, the integrated IMLR--
Fe abundance times gas mass per stellar mass--  is a steep function of radius,
which indicates
 that the gas is more extended than the stars in a cluster.
The  gas density profiles in the central regions of groups and poor
clusters were observed  to be shallower than those in the self-similar model,
and the relative entropy level was correspondingly higher than that
 in rich clusters \citep{Ponman99, Ponman03, Sun09, Pratt10}.
These deviations are considered to be best characterized by the injection
of energy (pre-heating) into the gas before the clusters collapse
(\cite{Kaiser91}).
Then, the difference in the ratio of gas mass to stellar mass may reflect
differences in distributions of gas and stars, which reflects history of 
energy injection from galaxies to the ICM.
To study the fractions of stars and gas in clusters of galaxies,  measurements 
of gas  and stellar mass beyond $r_{500}$ are necessary.

\subsection{ Ni/Fe ratio}

 The observed flat Fe abundance 
of the Coma cluster within 0.2 $r_{180}$ indicates that the
gas has been mixed well.
Since the Coma cluster does not have a   strong cool core ,
the derived  Ni/Fe ratio of the ICM is free from the resonant line scattering,
and the effect of metals recently ejected by cD galaxies is small.
In contrast,
the cool core of clusters comprise a mixture of metals present 
in the ICM and supplied from the cD galaxy, which in turn
contain metals that are synthesized by SN Ia and derived from stars
thorough stellar mass loss.

\begin{figure}[t]
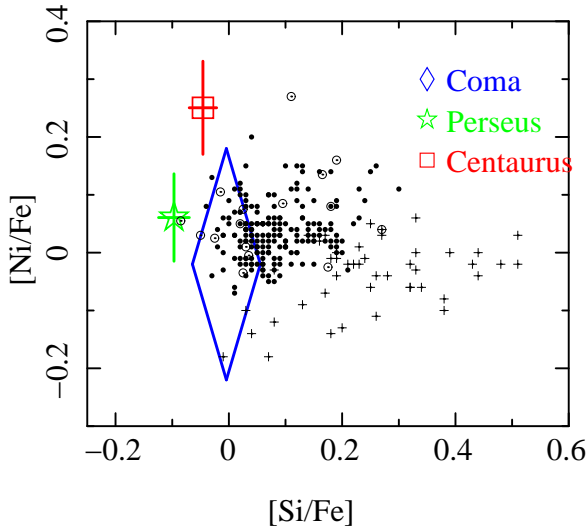

\FigureFile(80mm, 60mm){PASJ4137_figure7.ps}
\caption{
[Ni/Fe] of the central region of the Coma cluster (blue diamond),
within $\sim50$ kpc of the Perseus cluster (green star) and Centaurus
cluster (red square) are plotted against [Si/Fe].
Values of the Galactic stars from \citet{Edv1993}, \citet{feltzing1998},
and \citet{Gratton2003} are also plotted as filled circles, open circles,
and crosses, respectively.
}
\label{fig:nisife}
\end{figure}

The observed Ni/Fe ratio in the Coma cluster is 
  0.6--1.5 solar,   and
is consistent with those observed in the cool cores derived by \citet{dePlaa2007}.
Figure \ref{fig:nisife} shows the observed abundance ratios of [Ni/Fe] 
plotted against [Si/Fe] for 
the central region ($<$0.16 $r_{180}$) of the Coma cluster
and those in the cool cores of the Perseus and the Centaurus clusters 
\citep{Matsushita2011b}.
The figure also compares the abundance pattern with those  of the Galactic stars.
The stars with high [Si/Fe] reflect the average Si/Fe ratio synthesized
by SNe II in our Galaxy.
The [Si/Fe] of the three clusters are located around the stars with the smallest
[Si/Fe]. 
The [Ni/Fe] of the Coma cluster is consistent with 
those of the Galactic stars.
Therefore, for the ICM in the Coma cluster, the abundance pattern of 
Si, Fe, and Ni is consistent with the same mixture of the yields of  SN II and SN Ia
in our Galaxy, and an additional source of metals is not required.

\section{Conclusion}

We analyzed XMM (up to 0.5 $r_{180}$) and Suzaku 
(up to 0.16 $r_{180}$)
 data of the Coma cluster,  which is one of the
largest clusters in the nearby Universe.
Since the Coma cluster does not have   a  strong cool core   and the Fe abundance in the
ICM is flat up to 0.2 $r_{180}$, the derived abundance pattern is not affected
by recent metal supplies from the cD galaxies.

The Si/Fe ratios in the ICM are derived from the flux ratios of 
Ly$\alpha$ line of H-like Si and K$\alpha$ line of He-like Fe.
 The small temperature dependence of the line ratio limits
the systematic uncertainty in the derived abundance ratio.
The derived Si/Fe ratio in the ICM shows no radial gradient.
The emission weighted average of the Si/Fe ratio within 0.0--0.2 $r_{180}$,
0.2--0.5 $r_{180}$, and 0.0--0.5 $r_{180}$
is 0.97 $\pm$ 0.11, 1.05 $\pm$ 0.36, and 0.99 $\pm$ 0.13 in units of  solar ratio, using the solar abundance table by \citet{lodd2003}.
The comparison of the Si/Fe ratio of the Coma cluster with those of the smaller clusters
 indicates that dependence on the system mass of the Si/Fe ratio is small. 
The Ni/Fe ratio in the ICM is also derived from the flux ratio of the 7.8 keV line 
blend and K$\alpha$ line of He-like Fe.
The derived Ni/Fe ratio is 0.6--1.5 solar.
Therefore,  the abundance pattern of Si, Fe, and Ni is consistent with the same mixture of the yields of SN II and SN Ia in our Galaxy.

We calculated the cumulative IMLR up to 0.5 $r_{180}$ using K-band and B-band luminosities
of galaxies. Considering the observed Si/Fe ratio, at 0.5 $r_{180}$
the metal mass-to-light ratio in the ICM is consistent with expected value using a Salpeter
IMF. However, if the IMLR continues to increase with a radius beyond 0.5 $r_{180}$, 
 a flatter IMF is necessary.
The IMLR of the Coma cluster is similar to those of poor clusters with temperatures
of 2--4 keV. These clusters may have universal metal enrichment histories.


\begin{thebibliography}{}

\bibitem[Anders \& Grevesse(1989)]{angr} Anders, E., \& Grevesse, N.,\ 1989, \gca, 53, 197 

\bibitem[Arnaud et al.(2001)]{arnaud01} Arnaud, M., Aghanim, N., Gastaud, R., et al.\ 2001, \aap, 365, L67

\bibitem[Arnaud et al.(2007)]{Arnaud07} Arnaud, M., Pointecouteau, E., \& Pratt, G.~W.\ 2007, \aap, 474, L37 

\bibitem[Baumgartner et al.(2005)]{B2005} Baumgartner, W.~H., 
Loewenstein, M., Horner, D.~J., \& Mushotzky, R.~F.\ 2005, \apj, 620, 680 

\bibitem[B{\"o}hringer et al.(2004)]{Hans2004} B{\"o}hringer, H., Matsushita, K., Churazov, E., Finoguenov, A., \& Ikebe, Y.\ 2004, \aap, 416, L21 

\bibitem[Briel et al.(1992)]{briel1992} Briel, U. G., Henry, J. P. \& B{\"o}hringer, H.\ 1992, \aap, 259, L31


\bibitem[Churazov et al.(2004)]{Churazov2004} Churazov, E., Forman, 
W., Jones, C., Sunyaev, R., {\"o}hringer, H.\ 2004, \mnras, 347, 29 


\bibitem[de Grandi \& Molendi(2001)]{deGrandi2001} de Grandi, S., \& Molendi, S.\ 2001, \apj, 551, 153 


\bibitem[de Grandi \& Molendi(2002)]{deGrandi2002} de Grandi, S., \& Molendi, S.\ 2002, Chemical Enrichment of Intracluster and Intergalactic Medium, 253, 3 


\bibitem[de Grandi \& Molendi(2009)]{deGrandi2009} de Grandi, S., \& Molendi, S.\ 2009, \aap, 508, 565 

\bibitem[de Plaa et al.(2007)]{dePlaa2007} de Plaa, J., Werner, N., Bleeker, J.~A.~M., et al.\ 2007, \aap, 465, 345 


\bibitem[Doi et al.(1995)]{Doi} Doi, M., Fukugita, M., 
Okamura, S., \& Tarusawa, K.\ 1995, \apjs, 97, 77 

\bibitem[Edvardsson et  al.(1993)]{Edv1993} Edvardsson, B., Andersen, J., Gustafsson, B., et al.\ 1993, \aap, 275, 101 
\bibitem[Evrard et al.(1996)]{Evrard1996}
 Evrard,~A.~E., Metzler,~C.~A., \& Navarro,~J.~F.\
 1996, \apj, 469, 494 

\bibitem[Feltzing \& Gustafsson(1998)]{feltzing1998} Feltzing, S., \& Gustafsson, B.\ 1998, \aaps, 129, 237 

\bibitem[Finoguenov et al.(2000)]{Finoguenov2000}
Finoguenov, A., David, L. P. \& Ponman, T. J. 
2000, \apj, 544, 188
\bibitem[Finoguenov et al.(2001)]{Finoguenov2001}
Finoguenov, A., Arnaud, M. \& David, L. P. 
2001, \apj, 555, 191

\bibitem[Fukazawa et al.(1998)]{Fukazawa1998}
Fukazawa, Y., Makishima, K., Tamura, T., Ezawa, H., 
Xu, H., Ikebe, Y., Kikuchi, K. \& Ohashi, T. 1998, \pasj, 50, 187
\bibitem[Fukazawa et al.(2000)]{Fukazawa2000}
Fukazawa, Y., Makishima, K., Tamura, T., Nakazawa, K., 
Ezawa, H., Ikebe, Y., Kikuchi, K. \& Ohashi, T. 2000, \mnras, 313, 21

\bibitem[Gratton et al.(2003)]{Gratton2003} Gratton, R.~G., Carretta, E., Claudi, R., Lucatello, S., \& Barbieri, M.\ 2003, \aap, 404, 187 


\bibitem[Giodini et al.(2009)]{Gio09} Giodini, S., et al.\ 
2009, \apj, 703, 982 

\bibitem[Hayashi et al.(2009)]{Hayashi2009} Hayashi, K., Fukazawa, 
Y., Tozuka, M., et al.\ 2009, \pasj, 61, 1185 

\bibitem[Humphrey et al.(2006)]{Humphrey2006} Humphrey, P.~J., 
Buote, D.~A., Gastaldello, F., et al.\ 2006, \apj, 646, 899 


\bibitem[Iwamoto et al.(1999)]{Iwamoto1999} Iwamoto, K., Brachwitz, F., Nomoto, K., Kishimoto, N., Umeda, H., Hix, W.~R., \& Thielemann, F.-K.\ 1999, \apjs, 125, 439

\bibitem[Johnson et al.(2011)]{Johnson2011} Johnson, R., 
Finoguenov, A., Ponman, T.~J., Rasmussen, J., 
\& Sanderson, A.~J.~R.\ 2011, \mnras, 413, 2467 


\bibitem[Kaastra (1992)]{kaastra} Kaastra ~J.S. 1992, An X-Ray Spectral Code for Optically Thin Plasmas (Internal SRON-Leiden Report, updated version 2.0)


\bibitem[Kaiser(1991)]{Kaiser91}
 Kaiser,~N.\ 1991, \apj, 383, 104 

\bibitem[Komiyama et al.(2009)]{Komiyama2009}
 Komiyama, M., Sato, K., Nagino, R., Ohashi, T., \& Matsushita, K.\
 2009, \pasj, 61, S337 

\bibitem[Koyama et al.(2007)]{koyama07} Koyama, K., et al.\ 2007, \pasj, 59, 23 

\bibitem[Liedahl et al.(1995)]{liedahl} Liedahl ~D. A., Osterheld ~A.L., \& Goldstein ~W.H.  1995, ApJ, 438,     L115

\bibitem[Lin et al.(2003)]{Lin2003} Lin, Y.-T., Mohr, J.~J., 
\& Stanford, S.~A.\ 2003, \apj, 591, 749 

\bibitem[Lin et al.(2004)]{Lin2004} Lin, Y.-T., Mohr, J.~J., \& Stanford, S.~A.\ 2004, \apj, 610, 745 
\bibitem[Lodders(2003)]{lodd2003} Lodders, K.\ 2003, \apj, 591, 1220

\bibitem[Markevitch et al.(1998)]{Markevitch1998}
 Markevitch,~M., Forman,~W.~R., Sarazin,~C.~L., \& Vikhlinin,~A.\
 1998, \apj, 503, 77 
\bibitem[Makishima et al.(2001)]{Makishima2001}
Makishima,~K., et al.\ 2001, \pasj, 53, 401 
\bibitem[Matsushita et al.(2003)]{Matsushita2003}
Matsushita, K., Finoguenov, A. \& B{\"o}hringer, H. 2003, \aap, 401, 443

\bibitem[Matsushita et al.(2007a)]{Matsushita2007a} 
Matsushita,~K., et al.\ 2007a, \pasj, 59, 327 

\bibitem[Matsushita et al.(2007b)]{Matsushita2007}
Matsushita, K., B{\"o}hringer, H., Takahashi, I. \& 
Ikebe, Y., 2007b, \aap, 462, 953
\bibitem[Matsushita (2011)]{Matsushita2011} Matsushita, K.\ 2011, \aap, 527, A134

\bibitem[Matsushita \& Tamura (2011)]{Matsushita2011b} Matsushita, K. and Tamura, T., \ 2011, submitted to A\&A

\bibitem[Mewe et al.(1985)]{mewe85} Mewe ~R., Gronenschild ~E.H.B.M., \& van~den~Oord ~G.H.J. 1985, A\&AS, 62,197
\bibitem[Mewe et al.(1986)]{mewe86} Mewe ~R., Lemen ~J.R., \& van~den~Oord,~G.H.J. 1986, A\&AS,
        65,511


\bibitem[Michard \& Andreon(2008)]{Michard2008} Michard, R., \& Andreon, S.\ 2008, \aap, 490, 923 

\bibitem[Mitsuda et al.(2007)]{mitsuda07} Mitsuda, K., et al.\ 2007, \pasj, 59, 1 

\bibitem[Murakami et al.(2011)]{Murakami2011} Murakami, H., 
Komiyama, M., Matsushita, K., et al.\ 2011, \pasj, 63, 963 

\bibitem[Nagino \& Matsushita (2009)]{Nagino2009} Nagino, R., \& Matsushita, K.\ 2009, \aap, 501, 157 
\bibitem[Nomoto et al.(2006)]{Nomoto2006}
 Nomoto,~K., Tominaga,~N., Umeda,~H., Kobayashi,~C., 
\& Maeda,~K.\ 2006, Nuclear Physics A, 777, 424 

\bibitem[Ponman et al.(1999)]{Ponman99}
 Ponman,~T.~J., Cannon,~D.~B., \& Navarro,~J.~F.\ 1999, \nat, 397, 135 

\bibitem[Ponman et al.(2003)]{Ponman03}
 Ponman,~T.~J., Sanderson,~A.~J.~R., \& Finoguenov,~A.\
 2003, \mnras, 343, 331 


\bibitem[Pratt et al.(2010)]{Pratt10} Pratt, G.~W., et al.\ 2010, \aap, 511, A85 

\bibitem[Rasmussen \& Ponman(2007)]{Rasmussen2007}
 Rasmussen,~J., \& Ponman,~T.~J.\ 2007, \mnras, 380, 1554 

\bibitem[Rasmussen \& Ponman(2009)]{Rasmussen2009} Rasmussen, J., \& Ponman, T.~J.\ 2009, \mnras, 399, 239 
\bibitem[Renzini(2005)]{Renzini2005} 
Renzini, A.\ 2005, The Initial 
Mass Function 50 Years Later, 327, 221 

\bibitem[Sakuma et al.(2011)]{Sakuma2011} Sakuma, E., Ota, N., 
Sato, K., Sato, T., \& Matsushita, K.\ 2011, \pasj, 63, 979 


\bibitem[Sanders \& Fabian(2006)]{Sanders2006} Sanders, J.~S., \& Fabian, A.~C.\ 2006, \mnras, 371, 1483 

\bibitem[Sato et al.(2007a)]{Sato2007a}
Sato, K. \etal\ 2007a, \pasj, 59, 299
\bibitem[Sato et al.(2007b)]{Sato2007b}
Sato, K., Tokoi, K., Matsushita, K., Ishisaki, Y., Yamasaki, N. Y., Ishida, M. \& Ohashi, T. 2007b, \apj, 667, 41
\bibitem[Sato et al.(2008)]{Sato2008}
Sato, K., Matsushita, K., Ishisaki, Y., Yamasaki, N. Y., Ishida, M., Sasaki, S. \& Ohashi, T. 2008, \pasj, 60, 333

\bibitem[Sato et al.(2009a)]{Sato2009a} Sato, K., Matsushita, K., 
Ishisaki, Y., et al.\ 2009a, \pasj, 61, 353 
\bibitem[Sato et al.(2009b)]{Sato2009b}
Sato, K., Matsushita, K. \& Gastaldello, F., 2009b, \pasj, 61, 365

%
\bibitem[Sato et al.(2010)]{kSato2010} Sato, K., Kawaharada, M., 
Nakazawa, K., Matsushita, K., Ishisaki, Y., Yamasaki, N.~Y., 
\& Ohashi, T.\ 2010, \pasj, 62, 1445 



\bibitem[Sato et al.(2011)]{Sato2011} 
Sato, T., Matsushita, K., 
Ota, N., et al.\ 2011, \pasj, 63, 991 


\bibitem[Schlegel et al.(1998)]{Schlegel1998}
 Schlegel,~D.~J., Finkbeiner,~D.~P., \& Davis,~M.\ 1998, \apj, 500, 525 
%
\bibitem[Simionescu et al.(2009)]{Simionescu2009} Simionescu, A., Werner, N., B{\"o}hringer, H., et al.\ 2009, \aap, 493, 409 

\bibitem[Smith et al.(2001)]{Smith2001} 
 Smith,~R.~K., Brickhouse,~N.~S.,
 Liedahl,~D.~A., \& Raymond,~J.~C.\ 2001, \apjl, 556, L91 

\bibitem[Sun et al.(2009)]{Sun09} Sun, M., Voit, G.~M., 
Donahue, M., Jones, C., Forman, W., \& Vikhlinin, A.\ 2009, \apj, 693, 1142 

\bibitem[Tamura et al.(2004)]{Tamura2004} Tamura, T., Kaastra, J.~S., den Herder, J.~W.~A., Bleeker, J.~A.~M., \& Peterson, J.~R.\ 2004, \aap, 420, 135 


\bibitem[Tamura et al.(2009)]{tamura2009} Tamura, T., Maeda, Y., 
Mitsuda, K., et al.\ 2009, \apjl, 705, L62 

\bibitem[Tawara et al.(2008)]{Tawara2008} Tawara, Y., Matsumoto, 
C., Tozuka, M., et al.\ 2008, \pasj, 60, 307 
\bibitem[Tokoi et al.(2008)]{Tokoi2008} Tokoi,~K., et al.\ 2008, \pasj, 60, 317 

\bibitem[Vikhlinin et al.(1994)]{Vik94} Vikhlinin, A., 
Forman, W., \& Jones, C.\ 1994, \apj, 435, 162 
\bibitem[Vikhlinin et al.(2001)]{Vik01} Vikhlinin, A., 
Markevitch, M., Forman, W., \& Jones, C.\ 2001, \apjl, 555, L87 

\bibitem[Vikhlinin et al.(2006)]{Vik06} Vikhlinin, A., 
Kravtsov, A., Forman, W., Jones, C., Markevitch, M., Murray, S.~S., 
\& Van Speybroeck, L.\ 2006, \apj, 640, 691 

\bibitem[Werner et al.(2006)]{Werner2006} Werner, N., de Plaa, J., Kaastra, J.~S., et al.\ 2006, \aap, 449, 475 

\bibitem[Woosley \& Weaver(1995)]{Woos1995} Woosley, S.~E., \& Weaver, T.~A.\ 1995, \apjs, 101, 181 
\end{thebibliography}
\end{document}